\documentclass[12pt,reqno]{article}
\usepackage{amsmath,hyperref,amssymb}
\usepackage{graphicx,url}

\allowdisplaybreaks \numberwithin{equation}{section}

\DeclareSymbolFont{AMSa}{U}{msa}{m}{n}
\DeclareSymbolFont{AMSb}{U}{msb}{m}{n}
\DeclareMathSymbol{\fieldR}{\mathalpha}{AMSb}{"52}

\begin{document}

\begin{flushright} \small
ITP--UU--12/03 \\ SPIN--12/03
\end{flushright}
\bigskip
\begin{center}
 {\large\bfseries Black branes in AdS: BPS bounds and asymptotic charges}\\[5mm]
Kiril Hristov$^{*,\dag}$, Chiara Toldo$^*$, Stefan Vandoren$^*$ \\[3mm]
 {\small\slshape
 * Institute for Theoretical Physics \emph{and} Spinoza Institute, \\
 Utrecht University, 3508 TD Utrecht, The Netherlands \\
\medskip
 \dag Faculty of Physics, Sofia University, Sofia 1164, Bulgaria\\
\medskip
 {\upshape\ttfamily K.P.Hristov, C.Toldo, S.J.G.Vandoren@uu.nl}\\[3mm]}
\end{center}
\vspace{5mm} \hrule\bigskip \centerline{\bfseries Abstract}
\medskip
We focus on black branes and toroidal black holes in $N=2$ gauged supergravities that asymptote to AdS$_4$, and derive formulas for the mass and central charge densities. We derive the corresponding BPS bound from the superalgebra of the asymptotic vacuum and illustrate our procedure with explicit examples of genuine black brane solutions with non-trivial scalars.

\bigskip
\hrule\bigskip

\section{Introduction}\label{sect:intro}

BPS bounds for classical solutions in field and gravity theories are often derived from the superalgebra of the asymptotic form of the solution, at least when the solution can be embedded in a supersymmetric theory and its asymptotic form preserves some supersymmetry. We carried out this program and determined the BPS bounds for stationary black holes in AdS$_4$ for $N=2, D=4$ gauged supergravities in \cite{Hristov:2011ye,Hristov:2011qr}. In this paper we repeat the analysis for black branes and black holes with toroidal (instead of spherical) horizons.

While an event horizon with nonspherical topology is forbidden for asymptotically Minkowski solutions, in asymptotically AdS$_4$ spacetimes there exist solutions with negative and zero curvature. These objects share some of the thermodynamical properties of the positive curvature ones, like the entropy-area law. They can be formed after gravitational collapse \cite{Smith:1997wx} and present interesting properties in relation to the information loss paradox \cite{Klemm:1998bb}. Black branes are also important for various applications of the AdS/CFT correspondence, which partly motivates our study. The main outcome of our analysis is that for static black branes and toroidal black holes which are excitations above a supersymmetric vacuum, the BPS bound
\begin{align}\label{BPS-riads}
    M \geq |Z|\ ,
\end{align}
must be satisfied. Here the mass $M$ and the central charge $Z$ are explicit boundary integrals that will be derived in the coming sections. We illustrate the bound with some examples from the literature, describing in the process a general class of magnetic supersymmetric black branes based on the example provided by \cite{Cacciatori:2009iz} and the analysis of \cite{Hristov:2010ri}.

To apply the procedure of \cite{Hristov:2011ye} to black branes, we start with the Dirac brackets of two supercharges in gauged $N=2$ supergravity. Evaluated on shell, it is a surface integral on the boundary $\partial V$, and it reads
\begin{equation}\label{basic-susy-anticommutator-minimal}
\{\mathcal{Q},\mathcal{Q} \}  = 2 \oint_{\partial V} {\rm d}\Sigma_{\mu \nu}(\epsilon^{\mu \nu \rho \sigma}
\overline{\epsilon} \gamma_5 \gamma_{\rho} \widetilde{\mathcal{D}}_{\sigma} \epsilon)\ ,
\end{equation}
where $\widetilde{\mathcal{D}}_{\sigma} \epsilon$ is the gravitino supersymmetry variation (in our notation, $\epsilon$ contains two Majorana spinors). For supersymmetric solutions $\{\mathcal{Q},\mathcal{Q} \}$ vanishes, while for non-supersymmetric ones it is nonzero and positive. The BPS bound can then be determined from this relation by stripping off the Killing spinors and imposing positivity on the space of solutions, as shown explicitly in the following.

\section{Minimal gauged supergravity}\label{sect:toroidal}

We start from the minimal $D=4$ $N=2$ gauged supergravity, whose lagrangian and conventions can be found in \cite{Hristov:2011ye}, and consider static metrics of the form
\begin{equation}\label{formbrane}
{\rm d}s^2= U^2(r) \, {\rm d}t^2 -\frac{1}{U^2(r)} {\rm d}r^2 - r^2 {\rm d}\sigma^2\ , \qquad \qquad
U^2(r)= g^2 r^2 - \frac{2 \eta}{r} + \frac{q_e^2 + q_m^2}{r^2}\,,
\end{equation}
with a toroidal area element (with $\mathcal{V}$ the volume of the torus and $\tau$ the complex structure)
\begin{equation}\label{area-element}
{\rm d}\sigma^2 = \frac{\mathcal{V}}{{\rm Im}\tau} ({\rm d}x^2 + 2 {\rm Re}\tau \, {\rm d}x {\rm d}y + |\tau|^2 {\rm d}y^2) \,.
\end{equation}
The electromagnetic 1-form (the graviphoton) and its corresponding field strength are:
\begin{equation}\label{formbrane2}
A = \frac{q_e}{r} {\rm d}t + q_m \mathcal{V} x {\rm d}y \,, \qquad F = \frac{q_e}{r^2}\ {\rm d}t \wedge {\rm d}r + q_m \mathcal{V}\ {\rm d}x \wedge {\rm d}y\ .
\end{equation}
The range of the coordinates is restricted to $x \in [0,1]$, $y \in [0, 1]$ with periodic boundary conditions. The case of black branes can be obtained by decompactifying the torus, e.g. by considering a rectangular torus with ${\cal V}=R_1R_2, \tau=iR_2/R_1$ and sending the radii $R_1$ and $R_2$ to infinity. Doing so, we can use the volume as a regulator for black branes, and mass and charge densities will therefore be finite and well-defined.

The above metric describes a class of static charged toroidal black hole solutions with compact horizons. They asymptote at $r\rightarrow \infty$ to the vacuum configuration with $\eta=q_e=q_m=0$, which is a quotient of AdS$_4$, due to the identifications on $x$ and $y$. This spacetime is sometimes referred to in the literature as Riemann-anti-de Sitter (RiAdS) \cite{Vanzo:1997gw}. Supersymmetric toroidal solutions with magnetic charge do not exist, as shown in \cite{Cacciatori:2009iz,Caldarelli:1998hg} from the analysis of the integrability condition. The conditions to have a supersymmetric solution of the form \eqref{formbrane}-\eqref{formbrane2} that asymptotes to RiAdS are:
\begin{equation}
\eta=q_m=0\
\end{equation}
 for arbitrary electric charge.
Thus magnetic ground states as mAdS$_4$, considered in \cite{Hristov:2011ye} for spherical symmetry, do not  appear in the case of toroidal topology. In what follows we will therefore restrict to configurations with vanishing magnetic charge, $q_m=0$.

We now determine the Killing spinors of the ground state, $\eta=q_e=q_m=0$, since they will be inserted in \eqref{basic-susy-anticommutator-minimal} to derive a BPS bound. The Killing spinor equations are obtained by imposing that the supersymmetry variation of the gravitinos is zero (see \cite{Hristov:2011ye} and \cite{Caldarelli:1998hg} for details and notation):
\begin{equation}\label{supercovderivative}
\delta_{\epsilon}\psi_{\mu}=\widetilde{\mathcal{D}}_\mu \epsilon=
(\partial_{\mu} - \frac14 \omega_{\mu}^{ab} \gamma_{ab}-\frac{i}{2} g \gamma_{\mu} +
ig A_{\mu} \sigma^2 +\frac14 F_{\lambda \tau} \gamma^{\lambda \tau} \gamma_{\mu} \sigma^2) \epsilon = 0 \,.
\end{equation}

Choosing upper triangular vielbein $e_{\mu}^a$
\begin{equation}\label{vielbeins}
e_t^0= U\,, \quad e_r^1= \frac{1}{U}\,, \quad e_x^2= \frac{r \sqrt{ {\rm Im} \tau } \sqrt{\mathcal{V}}}{|\tau|} \,, \quad e_x^3 = \frac{r  \sqrt{\mathcal{V}}}{|\tau|} \frac{{\rm Re}\tau}{\sqrt{{\rm Im} \tau}}\,, \quad e_y^3= \frac{r |\tau|  \sqrt{\mathcal{V}}}{\sqrt{{\rm Im}\tau} }\ ,
\end{equation}
one can straightforwardly derive the (non-vanishing) components of the spin connection,
\begin{equation}\label{spinconn}
\omega_{t}^{01}= - U \partial_r U, \quad \omega_{x}^{12}= - \frac{\sqrt{ {\rm Im} \tau} \sqrt{\mathcal{V}}}{|\tau|} U\,, \quad \omega_{x}^{13}= - \frac{{\rm Re}\tau  \sqrt{\mathcal{V}}}{|\tau|\sqrt{ {\rm Im} \tau}} U\,, \quad \omega_{y}^{13}= -\frac{|\tau| \sqrt{\mathcal{V}}}{\sqrt{{\rm Im} \tau}} U\,.
\end{equation}
The Killing spinors can now be computed from \eqref{supercovderivative}, and the solution, with arbitrary constant spinors $\epsilon_0$, is
\begin{equation}\label{prefactor2}
 \epsilon = e^{\frac{i}{2} log(r)\gamma_1}\left (1+ \frac{ig}{2} \left[ x \frac{\sqrt{{\rm Im} \tau}  \sqrt{\mathcal{V}}}{|\tau|} \left(\gamma_2 +\frac{{\rm Re} \tau}{ {\rm Im} \tau} \gamma_3 \right) +y \frac{|\tau| \sqrt{\mathcal{V}}}{\sqrt{ {\rm Im} \tau}}\gamma_3 + gt \gamma_0 \right] \left(1-i \gamma_1 \right) \right) \epsilon_0\,\,.
\end{equation}
Without restriction on $\epsilon_0$, all eight supercharges are preserved, but \eqref{prefactor2} does not respect the identification of the coordinates $x,y$ on the torus. One can make the Killing spinors well-defined on the torus by imposing a projection on $\epsilon_0$, namely
\begin{equation}\label{projection-torus}
\epsilon_0= P\epsilon_0\ ,\qquad P\equiv \frac{1+i\gamma_1}{2}\ .
\end{equation}
Such a projection breaks half of the supersymmetries, and the Killing spinors of RiAdS are therefore
\begin{equation}\label{spinortorus}
\epsilon_{RiAdS}  = \sqrt{r} \left( \frac{1+i\gamma_1}{2} \right) \epsilon_0 = \sqrt{r}P \epsilon_0 \,.
\end{equation}
This result agrees with the one found in \cite{Caldarelli:1998hg}. The toroidal black holes can be seen as excitations over the background characterized by these Killing spinors. To find the BPS bound, we plug the spinors \eqref{spinortorus} in the formula for the Dirac bracket of two supercharges \eqref{basic-susy-anticommutator-minimal}. We first note the relations involving the projector \eqref{projection-torus}:
\begin{align}\label{property1}
\begin{split}
 P i \gamma_1 P  =  P\ , \qquad    P \gamma_{02} P =  P ( -i\gamma_{012}) P = \gamma_{02} P\ ,\\  P \gamma_{03} P= P (-i\gamma_{013}) P = \gamma_{03} P \ , \qquad
P \gamma_{23} P= P i \gamma_{1 2 3} P =  \gamma_{23} P\ .
\end{split}
\end{align}
All the other gamma matrices between two projectors give zero: this strongly limits the number of terms present in the superalgebra. The anticommutator between two supercharges can now be computed. Due to the projection identities \eqref{property1} and the symmetries of the gamma matrices\footnote{$\gamma^{02},\gamma^{03}$ are symmetric in their spinor indices, while $\gamma^{23}$ is antisymmetric (see e.g.\ \cite{Hristov:2010ri} for our gamma matrix conventions). The four terms in \eqref{susy_commutator} are therefore the only non-vanishing contributions from \eqref{basic-susy-anticommutator-minimal}.} only four terms appear, and the result is
\begin{equation}\label{susy_commutator}
\{Q,Q\}= 2 \overline{(P \epsilon_0)}(- i M \gamma^0 - i P_2 \gamma^{2} - i P_3 \gamma^{3} - Z \gamma^{5} \sigma^2) P \epsilon_0 \,.
\end{equation}
 The mass $M$ has the following expression:
\begin{equation}\label{mass_min}
M= \frac12 \lim_{r\to \infty}{\oint {\rm d} \Sigma_{tr} e_{0}^{t} e_{1}^{r} \left(\, 2 g r -r (\omega_{x}^{1 2} e_{2}^{x} + \omega_{y}^{13} e_{3}^{y} + \omega_{y}^{12} e_{2}^{y} ) \,  \right)}\ ,
\end{equation}
and the central charge $Z$ reads:
\begin{equation}\label{centralcharge}
Z=  \lim_{r\to \infty}\oint_{T^2} r F\ ,
\end{equation}
where $F$ is the vector field strength written as a two-form. The above formulas are valid after choosing an upper triangular vielbein, as in \eqref{vielbeins}. We omit the formulas for the momenta $P_2$ and $P_3$, which are straightforward to derive from \eqref{basic-susy-anticommutator-minimal} and \eqref{spinortorus}, but are not particularly insightful since they vanish for static solutions.

The full superalgebra of RiAdS can be then most clearly presented as follows. After the projection we have only 4 real supercharges present, which we label  $Q_1,Q_2,Q_3,Q_4$. The non-vanishing supercharge anticommutators can then be read from \eqref{susy_commutator}:
\begin{align}
\begin{split}
\{Q_1, Q_1 \} &= \{Q_3, Q_3 \}= M + P_2\,, \qquad \{Q_2, Q_2 \} = \{Q_4, Q_4 \} = M - P_2\,,\\
 \{Q_1, Q_2 \} &=  \{Q_3, Q_4 \} = P_3\,, \qquad \{Q_1, Q_4 \} = - Z\ , \qquad \{Q_2, Q_3 \} = Z\ ,
\end{split}
\end{align}
 and the action of the gauged $U(1)_R$ symmetry leads to
\begin{equation}
 [Q_1,T]= Q_3\,, \quad [Q_2,T]= Q_4\ , \quad  [Q_3,T]= -Q_1\,, \quad [Q_4,T]= -Q_2\ .
\end{equation}
The other commutators vanish due to the form of the Killing vectors, the fact that gauge transformations commute with translations and compatibility with the super Jacobi identities. This shows that indeed $Z$ is a central charge, similar to the magnetic central charge in the Poincar\'{e} superalgebra.

Due to the toroidal compactification, the theory is endowed also with modular invariance. The metric and the Killing spinor are invariant under transformations that act on both the parameter $\tau$ and the coordinates $(x,y)$:
\begin{equation}\label{transfmodular}
\tau \rightarrow  \frac{a \tau +b}{ c \tau +d} \,, \qquad
\left( \begin{array}{c} x \\
y \\
\end{array} \right) \rightarrow
M
\left( \begin{array}{c} x \\
y \\
\end{array} \right) =
\left( \begin{array}{cc} a & b \\
c & d \\
\end{array} \right)
\left( \begin{array}{c} x \\
y \\
\end{array} \right)\,,
\end{equation}
with the condition $ad-bc=1$, i.e. $M \in SL(2, Z)$. For a finite volume of the torus, the superalgebra can be interpreted as corresponding to a modular invariant quantum mechanics theory in one dimension, since one can further reduce the $3d$ theory, dual to RiAdS, on the two compact spatial dimensions. This interpretation is no longer valid in the infinite volume limit where the boundary is an infinite flat plane.

To give an explicit example how the BPS bound constrains the solutions space, we now restrict our attention to static solutions of the form \eqref{formbrane} with zero magnetic charge. The above mass formula can be explicitly evaluated:
\begin{align}
\begin{split}
M & =  \frac{1}{2} \lim_{r\to \infty} {\oint {\rm d}x {\rm d}y \mathcal{V} \,r^2 \,\left[ 2 g r +r \left(\omega_x^{12} \frac{|\tau|}{r \sqrt{ {\rm Im} \tau}} + \omega_y^{13} \frac{\sqrt{{\rm Im}\tau}}{r |\tau|} \right) \right]} = \\
& =  \lim_{r\to \infty} {\oint {\rm d}x {\rm d}y \mathcal{V} \, \left( gr^3 - r^2 \sqrt{r^2 g^2 -\frac{2 \eta}{r} + \frac{q_e^2}{r^2}} \right)=  \mathcal{V} \int_0^1 {\rm d}x \int_0^1{\rm d}y \, \, \eta} = \mathcal{V}\, \eta \,.
\end{split}
\end{align}
We see that the divergent part cancels (notice that the dependence on $\tau$ drops out, as a consequence of the modular symmetry) and we are left with the finite quantity $\eta$ for the mass density $M/\mathcal{V}$. Furthermore, the formula for the central charge \eqref{centralcharge} gives zero when computed on the ansatz \eqref{formbrane},
\begin{equation}\label{cencharge}
Z = q_m r \mathcal{V} =  0\ ,
\end{equation}
since $q_m$ is forced to vanish for asymptotically RiAdS solutions. We then have $P_2=P_3=Z =0, M = \eta \mathcal{V}$ for these solutions. Consequently, the BPS bound is just:
\begin{equation}
  \eta \geq 0 \,.
\end{equation}
 Note that the BPS bound does not involve the electric charge. Moreover, this bound also holds in the decompactification limit for black branes, where the mass density $\eta$ is a finite number even if the mass $M$ is infinite. The BPS bound is saturated for $\eta=0$ with an arbitrary electric charge $q_e$. The resulting spacetime has a naked singularity whenever $q_e \neq 0$, which is often considered unphysical. In minimal gauged supergravity  there is therefore no genuine BPS black brane solution. To make the situation more appealing, we now turn to general gauged supergravities. We will see that turning on matter couplings allows us to generate non-zero central charge and mass for the BPS configurations, which also leads to the existence of supersymmetric black brane solutions with horizon.

\section{General gaugings}

As shown in \cite{Hristov:2011qr}, the generalization of the results from minimal supergravity to one with arbitrary vector and hypermultiplet gaugings is fairly straightforward. For simplicity, we will restrict the discussion here only to abelian gaugings. The superalgebra remains the same with the only difference that the definition of the asymptotic charges generalizes to accommodate for the possibility of non-constant scalars, see (1.5)-(1.6) of \cite{Hristov:2011qr}. For a solution with constant scalars (both in the vector and in the hypermultiplet sector) our results therefore remain as in the minimal case, as can be easily checked in the explicit expressions that follow.

We are mostly interested in describing objects with vanishing $P_2,P_3$ like static black holes and branes. The relevant asymptotic charges in this case are the mass $M$ and the central charge $Z$. In the general case with arbitrary vector and hypermultiplets, they are defined as (c.f. \cite{Hristov:2011qr} for details about notation):
\begin{equation}\label{mass}
M= \frac12 \lim_{r\to \infty}{\oint {\rm d} \Sigma_{tr} e_{0}^{t} e_{1}^{r} \left(\, 2 g r |P^a_{\Lambda} L^{\Lambda}| -r (\omega_{x}^{1 2} e_{2}^{x} + \omega_{y}^{13} e_{3}^{y} + \omega_{y}^{12} e_{2}^{y} ) \,  \right)}\ ,
\end{equation}
and
\begin{equation}\label{ccharge}
Z= 2 \lim_{r \rightarrow \infty} \oint_{T^2} r\ {\rm Im} \left( T^- \right) = 2 \lim_{r \rightarrow \infty} r \mathcal{V}\ {\rm Im}\left( L^{\Lambda} q_{\Lambda} - M_{\Lambda} p^{\Lambda} \right)\ ,
\end{equation}
where $T^-$ is the anti-selfdual part of the graviphoton field strength.

Compared to \eqref{mass_min}, the expression for the mass with vector and hypermultiplets is changed only slightly in order to accommodate for the cosmological constant, which is now dependent on the scalar fields via the expression $P^a_{\Lambda} L^{\Lambda}$. $P^a_{\Lambda}$ are the hypermultiplet moment maps, which can also be non-zero constants (Fayet-Iliopoulos (FI) parameters) in the absence of hypermultiplets, while $L^{\Lambda}$ (together with $M_{\Lambda}$) are the special geometry sections that depend on the complex scalars $z^i$ in the vector multiplets.

The expression for the central charge is reminiscent of the expression for the magnetic charge caused by the graviphoton field strength\footnote{$q_{\Lambda}, p^{\Lambda}$ are the electric and magnetic charge densities of the vector field strengths appearing in the lagrangian (see \cite{Hristov:2011qr} for details).}, just as in \eqref{cencharge}. The magnetic charge, $\lim_{r \rightarrow \infty} {\rm Im}\left( L^{\Lambda} q_{\Lambda} - M_{\Lambda} p^{\Lambda} \right)$, is forced to vanish due to supersymmetry of the vacuum as proven in \cite{Cacciatori:2009iz}. $Z$ is in fact the first subleading term in the expression for the magnetic charge due to the extra $r$ factor and is automatically finite in the limit $r \rightarrow \infty$. For constant scalars, \eqref{ccharge} clearly reduces to \eqref{cencharge} and the central charge vanishes. For non-constant scalar profiles, however, it is now possible to generate a non-zero $Z$, which turns out to be crucial for generating a massive BPS black brane with an event horizon.

The BPS bound for static asymptotically RiAdS solutions when a central charge is allowed is therefore
\begin{align}\label{BPS-riads1}
    M \geq |Z|\ ,
\end{align}
as already predicted. In case when both $M$ and $Z$ vanish we recover a 1/2 BPS solution like the ones in the previous section, while in case $M = |Z| \neq 0$ we have a 1/4 BPS excitation. All other cases result in non-supersymmetric excitations over RiAdS.

\subsection{Magnetic BPS black branes}
A class of BPS solutions with genuine horizons, corresponding to black branes and toroidal black holes in abelian gauged supergravity with FI terms, can be derived from the example in \cite{Cacciatori:2009iz} and following the steps in \cite{Hristov:2010ri}. The solutions are in complete analogy to the ones found in \cite{Hristov:2010ri} with the only exception that the flat horizon forces the magnetic charge carried by the graviphoton to vanish, $\xi_{\Lambda} p^{\Lambda} = 0$, as already mentioned above\footnote{The solution holds in gauged supergravity with FI parameters $P^a_{\Lambda} = \xi_{\Lambda} = const$. See \cite{Hristov:2010ri} for all technical details.}. In short, one can find a class of 1/4 BPS solutions, given by
 \begin{align}\label{our_solution}
 \begin{split}
 {\rm d} s^2 = e^{\mathcal{K}} \left(g r+\frac{c}{2 g r} \right)^2 {\rm d}t^2 - \frac{e^{-\mathcal{K}} {\rm d}r^2}{\left(g r+\frac{c}{2 g r} \right)^2} - e^{-\mathcal{K}} r^2 {\rm d} \sigma^2\ ,\\
 {\rm Re} X^{\Lambda} = H^{\Lambda} = \alpha^{\Lambda} + \frac{\beta^{\Lambda}}{r}, \qquad \qquad {\rm Re} F_{\Lambda} = 0\ ,\\
 \xi_{\Lambda} \alpha^{\Lambda} = - 1\ , \qquad \xi_{\Lambda}
  \beta^{\Lambda} = 0\ , \qquad F_{\Lambda} \left( -2 g^2 r \beta^{\Lambda} + c
  \alpha^{\Lambda}+2 g p^{\Lambda}\right)= 0\ ,
 \end{split}
 \end{align}
under the restriction $\xi_{\Lambda} p^{\Lambda} = 0$, with the toroidal area element given by \eqref{area-element}. These solutions satisfy the BPS bound $M = |Z|$, where both asymptotic charges are non-vanishing. Thus, unlike their spherical analogs (c.f. the discussion in section 4 of \cite{Hristov:2011qr}), the magnetic black branes have a non-vanishing mass.

To see this in some detail, consider the simple case of a prepotential $F =-2 i \sqrt{X^0 (X^1)^3}$. We have $X^0 = H^0 = \alpha^0+\frac{\beta^0}{r}, X^1 = H^1 = \alpha^1+\frac{\beta^1}{r}$ and $e^{-\mathcal{K}} = 8 \sqrt{H^0 (H^1)^3}$, with
 \begin{equation}\label{constants}
\beta^0 = -\frac{\xi_1 \beta^1}{\xi_0}, \qquad \alpha^0 = -\frac{1}{4 \xi_0}, \qquad \alpha^1 = -\frac{3}{4 \xi_1}, \qquad c = - \frac{32}{3} (g \xi_1 \beta^1)^2\ ,
\end{equation}
and magnetic charges
\begin{equation}\label{magn-charges}
p^0 = \frac{8 (g \xi_1 \beta^1)^2}{3 g \xi_0}, \quad p^1 = -\frac{8 (g \xi_1 \beta^1)^2}{3 g \xi_1}\ .
\end{equation}
Note that this solution is in almost complete analogy to the one discussed in section 7.1 of \cite{Hristov:2010ri} in the spherical case. It has a double horizon at $r_h = \frac{4}{\sqrt{3}} \xi_1 \beta^1$, which shields the singularity for any positive value of the arbitrary parameter $\xi_1 \beta^1$. Just as in the spherical example (section 4.2 of \cite{Hristov:2011qr}), we have to rescale the radial coordinate $r$ with $a = \lim_{r \rightarrow \infty} e^{-\mathcal{K}/2}$ in order to have the proper asymptotics. Evaluating \eqref{mass} and \eqref{ccharge} eventually leads to:
\begin{align}
M/\mathcal{V} = \lim_{r\rightarrow\infty} \frac{e^{-\mathcal{K}/2}}{a^2} r^2 \left(g r- e^{\mathcal{K}/2} \left( g r+\frac{a^2 c}{2 g r}  \right) \partial_r (r e^{-\mathcal{K}/2}) \right)
 = \frac{128}{9} g (\xi_1 \beta^1)^3\ ,
\end{align}
\begin{align}
Z/\mathcal{V} = \lim_{r\rightarrow\infty} 2 r e^{\mathcal{K}/2} \sqrt{\frac{H^1}{H^0}} (p^0 H^1 + 3 p^1 H^0) = \frac{128}{9} g (\xi_1 \beta^1)^3\ .
\end{align}
This proves that the mass is equal to the central charge. The solution is a 1/4 BPS toroidal black hole in RiAdS for any finite value of $\mathcal{V}$ and black brane in AdS as $\mathcal{V} \to \infty$. From the form of the superalgebra it is clear that one should in principle be able to add arbitrary electric charges to these solutions and still keep them supersymmetric. To our best knowledge, such solutions have not been yet constructed (see however \cite{Charmousis:2010zz,Barisch:2011ui} for supersymmetric and extremal electric black branes that do not strictly asymptote to AdS).

\section*{Acknowledgements}
We thank D. Klemm for helpful correspondence. We acknowledge support by the Netherlands Organization
for Scientific Research (NWO) under the VICI grant 680-47-603.


\begin{thebibliography}{00}


\bibitem{Hristov:2011ye}
  K.~Hristov, C.~Toldo and S.~Vandoren,
JHEP {\bf 1112}, 014 (2011),
  arXiv:1110.2688 [hep-th].

\bibitem{Hristov:2011qr}
  K.~Hristov,
  arXiv:1112.4289 [hep-th].


\bibitem{Smith:1997wx}
R.B. Mann and W.L. Smith, Phys. Rev. D {\bf 56}, 4942 (1997), arXiv:gr-qc/9703007.

 \bibitem{Klemm:1998bb}
 D. Klemm and L. Vanzo, Phys. Rev. D {\bf 58}, 104025 (1998), arXiv:gr-qc/9803061.

 \bibitem{Cacciatori:2009iz}
S.~Cacciatori and D.~Klemm,
JHEP {\bf 1001}, 085 (2010), arXiv:0911.4926 [hep-th].

\bibitem{Hristov:2010ri}
  K.~Hristov and S.~Vandoren,
  JHEP {\bf 1104}, 047 (2011),
  arXiv:1012.4314 [hep-th].

\bibitem{Vanzo:1997gw}
  L.~Vanzo, Phys.\ Rev.\ D\ {\bf 56}, 6475 (1997),  arXiv:gr-qc/9705004.

\bibitem{Caldarelli:1998hg}
M. Caldarelli and D. Klemm,
Nucl.\ Phys.\  B {\bf 545}, 434
(1999), arXiv:hep-th/9808097.


\bibitem{Charmousis:2010zz}
C. Charmousis, B. Gouteraux, B. S. Kim, E. Kiritsis and R. Meyer, JHEP {\bf 1011}, 151 (2011), arXiv:1005.4690 [hep-th].

  \bibitem{Barisch:2011ui}
  S.~Barisch, G.~L.~Cardoso, M.~Haack, S.~Nampuri and N.~A.~Obers, JHEP {\bf 1111} , 090 (2011),
 arXiv:1108.0296 [hep-th].

\end{thebibliography}
\end{document}